\documentclass{sig-alternate}
\usepackage{times}
\usepackage{verbatim}
\usepackage{multirow} %
\usepackage[numbers]{natbib}
\usepackage{amsmath}
\usepackage{amssymb}
\usepackage{array}
\usepackage{booktabs}
\usepackage{afterpage}
\usepackage[dvipsnames]{color}
\usepackage{mfirstuc}
\usepackage{subcaption}
\usepackage{multirow}
\usepackage{rotating}
\usepackage{listings}
\PassOptionsToPackage{pdfpagelabels=false}{hyperref}\usepackage{hyperref}
\usepackage{xspace}
\usepackage{tikz}
\usetikzlibrary{snakes,decorations.pathreplacing,calc}
\usepackage{CJKutf8}
\usepackage{xparse}

\newsavebox{\fminipagebox}
\NewDocumentEnvironment{fminipage}{m O{\fboxsep}}
 {\par\kern#2\noindent\begin{lrbox}{\fminipagebox}
  \begin{minipage}{#1}\ignorespaces}
 {\end{minipage}\end{lrbox}%
  \makebox[#1]{%
    \kern\dimexpr-\fboxsep-\fboxrule\relax
    \fbox{\usebox{\fminipagebox}}%
    \kern\dimexpr-\fboxsep-\fboxrule\relax
  }\par\kern#2
 }

\newcommand{\hide}[1]{} %
\newcommand{\para}[1]{\vspace{0.05in}\noindent\textbf{#1\xspace}}
\newcommand{\figref}[1]{Figure~\ref{#1}\xspace} %
\newcommand{\shortfigref}[1]{Fig.~\ref{#1}\xspace} %

\newdef{definition}{Definition}
\newdef{problem}{Problem}
\newcommand{\absolute}{fixed-prefix\xspace}
\newcommand{\Absolute}{Fixed-prefix\xspace}
\newcommand{\relative}{full-life\xspace}
\newcommand{\Relative}{Full-life\xspace}
\newcommand{\level}{quartile\xspace}
\newcommand{\levels}{quartiles\xspace}
\newcommand{\high}{the top\xspace}  %
\newcommand{\higher}{the higher-activity\xspace} %
\newcommand{\floornum}[1]{\lfloor #1 \rfloor}
\newcommand{\numWindows}{T_w} %

\newcommand{\subfigureplot}[4]{%
\begin{subfigure}[t]{#4\textwidth}
  \includegraphics[width=\textwidth]{#1}
  \caption{#2}
  \label{#3}
\end{subfigure}}
\newcommand{\unk}{\mbox{<RARE>}\xspace}
\newcommand{\major}{\mbox{50+}\xspace}
\newcommand{\words}{{\it words}\xspace} %
\newcommand{\staying}{lasting\xspace} %

\newif{\ifhidecomments}
\hidecommentsfalse   %

\ifhidecomments
   \newcommand{\llee}[1]{}
   \newcommand{\chenhao}[1]{}
\else
  \newcommand{\llee}[1]{\textcolor{red}{#1}}
  \newcommand{\chenhao}[1]{\textcolor{blue}{#1}}
\fi

\permission{Copyright is held by the International World Wide Web Conference Committee (IW3C2). IW3C2 reserves the right to provide a hyperlink to the author's site if the Material is used in electronic media.}
\conferenceinfo{WWW 2015,}{May 18--22, 2015, Florence, Italy.}
\copyrightetc{ACM \the\acmcopyr}
\crdata{978-1-4503-3469-3/15/05. \\
http://dx.doi.org/10.1145/2736277.2741661%
}
\begin{document}

\title{
All Who Wander: On the Prevalence and Characteristics of Multi-community Engagement
}

\numberofauthors{2}
\author{
\alignauthor
Chenhao Tan\\
\affaddr{Dept. of Computer Science}\\
\affaddr{Cornell University}\\
\email{chenhao@cs.cornell.edu}
\alignauthor
Lillian Lee\\
\affaddr{Dept. of Computer Science}\\
\affaddr{Cornell University}\\
\email{llee@cs.cornell.edu}
}

\maketitle
\begin{abstract}

Although
 analyzing user behavior \emph{within} individual communities
is an active and rich research domain,
people usually interact with \emph{multiple} communities both on- and off-line.
How do users act in such multi-community environments?
Although there are a host of intriguing aspects to this question,
it has received much less attention in the research community
in comparison
to the intra-community case.
In this paper, we examine three aspects of multi-community engagement:
the \emph{sequence of communities} that users post to,
the \emph{language} that users employ
in those communities,
and the 
\emph{feedback} that users receive,
using longitudinal posting behavior on Reddit as our main data source, and
DBLP for 
auxiliary experiments.
We also demonstrate the effectiveness of features drawn from these 
aspects
in predicting users' future level of activity.
One might expect that a user's trajectory mimics the ``settling-down'' process in real life:
an initial
exploration of 
sub-communities before settling down into a few
niches. However, we find that the users in our data 
continually post in new
communities;
moreover,
as time goes on,
they post %
increasingly
evenly among a more diverse set of smaller communities.
Interestingly, it seems that users that eventually leave the community are ``destined'' to do so from the very beginning, 
in the sense of showing
significantly different ``wandering'' patterns very early on in their trajectories;
this finding has potentially important design implications for community maintainers.
Our multi-community perspective also allows us to investigate the ``situation vs. personality'' debate from language usage across different communities.

\end{abstract}

\vspace{1mm}
\noindent
{\bf Categories and Subject Descriptors:} 
J.4 [{\bf Computer Applications}]: SOCIAL AND BEHAVIORAL SCIENCES;
H.2.8 [{\bf Database Applications}]: Data Mining

\vspace{1mm}
\noindent
{\bf General Terms:} 
Algorithms, Experimentation

\vspace{1mm}
\noindent
{\bf Keywords:}
multiple communities,
lifecycle,
language, Reddit, DBLP

\sloppy

\section{Introduction}
\label{sec:intro}

\hfill
\begin{CJK}{UTF8}{gbsn}
树挪死，人挪活
\end{CJK}
({\em People, unlike trees, thrive on relocation}).

\hfill ---A Chinese saying

\smallskip

How people behave {\em within} a given community is a
profound and broad
question
that
has inspired work ranging from basic social-science research (e.g., \cite{Shaw:1971a})
to the design of online social systems (e.g., \cite{evidence-based-community}).
However, many settings offer an array of {\em multiple} possible
interest sub-groups for users to engage in. In the offline world, for example,
within the bounds of a single college campus, students can get involved with a
variety of clubs, organizations, and social circles.
And in the online case, there are many multi-community sites, such as Reddit, 4chan, Wikia, and StackExchange, all of which host a slew of topic-based sub-discussion forums.
As the results in this paper show, multi-community settings exhibit many interesting and useful properties that are not manifested in within-community situations, and so
{\em our main goal is to demonstrate that multi-community engagement is an exciting and underexploited research area}: we believe that such work will shed additional light on human behavior and on the design of social-media systems.

To demonstrate, we first tackle a seemingly foregone conclusion:
that, analogously to
the
human life course \cite{Buhler:JournalOfAppliedPsychology:1935,erikson1998life},
a person first passes through
an
``adolescent'' phase of trying
out many different interests before ``settling down''.
Indeed, the best-paper award at WWW 2013 was given to an excellent within-community study \cite{Danescu-Niculescu-Mizil:2013:NCO:2488388.2488416} demonstrating (among other things) that users' language use becomes more inflexible and out-of-step with the community's over time.
But,  contrary to this expectation, we find that
even
people with long
histories of participation in a global community \emph{continually} try out new
sub-communities.
\figref{fig:cumu_sub_count} depicts this for two very different settings: for Reddit and for the universe of computer-science conferences
given by DBLP, the latter choice inspired by \cite{Backstrom:2006:GFL:1150402.1150412}.
Note that despite their very different timescales (one can post to Reddit at any time, but submission deadlines only
roll around every so often) and  barriers to entry (conferences have gate-keepers, whereas posting on
Reddit can be done essentially at will),
they exhibit the same qualitative behavior.
On average, Redditors
post to 5 communities in their first 10 posts and then post to 2.5 \emph{new} communities every 10 posts, while
researchers publish at 5 \emph{new} conferences every 10 papers (\shortfigref{fig:cumu_sub_count_reddit} and \ref{fig:cumu_sub_count_dblp}).
These exploration trends continue over the users' lifetimes
(\shortfigref{fig:cumu_rel_sub_count_reddit},  \ref{fig:cumu_rel_sub_count_dblp}).
Thus, while within a single community ``all users die old'' \cite{Danescu-Niculescu-Mizil:2013:NCO:2488388.2488416},
it seems that a multi-community setting keeps users young by offering them
choices to explore
as an alternative to opting out entirely.

\newcommand{\newr}[1]{\underline{#1}}
\begin{figure*}[t]
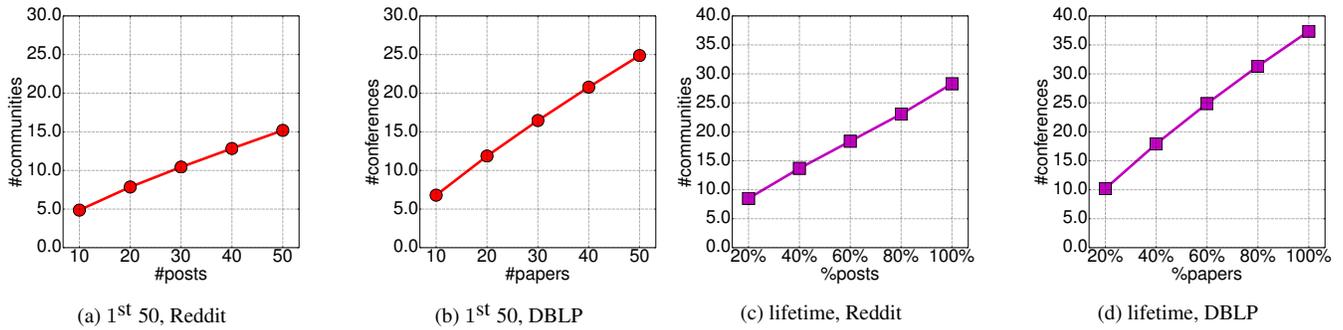

\centering
\subfigureplot{{plots/1107/cumu_sub_count/absolute_all_50_10_10_5}.pdf}
{$1^{\mbox{st}}$ 50, Reddit}{fig:cumu_sub_count_reddit}{0.23}
\hfill
\subfigureplot{{plots/1012/dblp/count/cumu_absolute_all_50_10_10_5}.pdf}
{$1^{\mbox{st}}$ 50, DBLP}{fig:cumu_sub_count_dblp}{0.23}
\subfigureplot{{plots/1107/cumu_sub_count/relative_all_50_10_10_5}.pdf}{
lifetime, Reddit}{fig:cumu_rel_sub_count_reddit}{0.23}
\hfill
\subfigureplot{{plots/1012/dblp/count/cumu_relative_all_50_10_10_5}.pdf}{
lifetime, DBLP}{fig:cumu_rel_sub_count_dblp}{0.23}
\caption{
Mean number of {\bf unique} communities (subreddits for Reddit, conferences for
DBLP) where people make their
temporally
first $x$ contributions (left-hand plots) or their first $x$
{percent} of contributions (right-hand plots), for ``long-lived'' people (50+
contributions overall).  For Reddit (respectively, DBLP), contributions = posts (papers).
Standard-error intervals are depicted, but very small, and trends for the median are consistent with the mean.
Note that the left-hand plots depict long timespans: the average time
to accumulate 50 contributions is
456.0
days on Reddit, 15.6 years on DBLP.
\smallskip \\
{\small
{\em Example of a Redditor's first 50 subreddits, in the order
posted to, first-time communities underlined}:
\newr{skyrim}, \newr{aww}, skyrim, aww, \newr{pics}, aww, aww, pics, \newr{WTF}, aww, pics, WTF, \newr{pokemontrades}, \newr{funny}, pokemontrades, pics, aww, \newr{AskReddit}, pics, \newr{pokemon}, \newr{fashion}, AskReddit, aww, \newr{Scotland}, fashion, aww, Scotland, pics, \newr{keto}, keto, \newr{Fitness}, keto, skyrim, pokemon, \newr{cats}, aww, aww, pokemon, Scotland, AskReddit, fashion, keto, pokemon, \newr{ketouk}, Scotland, keto, pics, ketouk, funny, \newr{gamecollecting}.
\\  {\em Two DBLP examples}:  the set of venues of \emph{James Harland}'s first 50 papers: LPAR, ACE, NACLP, TABLEAUX, DALT, ECOWS, CADE, Australian Joint Conference on Artificial Intelligence, IAT, ICLP, ICSOC, ILPS, ``Workshop on Programming with Logic Databases (Book), ILPS'', Future Databases, AAMAS, ACSE, EDBT, JICSLP, ACSC, ACAL, SAC, AAMAS (1), PRICAI, Computational Logic, CLIMA, ECAI, AMAST, ISLP, ``Workshop on Programming with Logic Databases (Informal Proceedings), ILPS'', KR, CATS.
\\
\emph{Jure Leskovec}'s: INFOCOM, HT, AAAI, PKDD, ICDE, ECCV (4), KDD, ICDM, UAI, NIPS, ICML, CHI, VLDB, WWW, EC, WAW, WSDM, ICWSM, PAKDD, CIKM-CNIKM, JCDL, SDM, WWW (Companion Volume).
} %
\label{fig:cumu_sub_count}}
\end{figure*}

Having established the prevalence of ``wandering'' behavior,
we are led to
investigate a host of related phenomena.  \emph{
We believe
that these phenomena are
interesting in their own right, and at times quite surprising. Moreover, we also demonstrate that our findings inspire
new kinds of features that are strongly predictive of users' future level of
activity.}

\para{Organization,
further paper highlights and design implications.}
In Sections \ref{sec:data} and \ref{sec:exploration}, we investigate
three 
aspects
of users' {community trajectories}:
the communities
they post to (\S \ref{sec:community}), the language
they use within a community (\S \ref{sec:languse}), and the 
feedback
they
receive from 
other members of the community (\S \ref{sec:feedback}).
Consistently, we see that --- again, in contrast to the ``older people become
less adventurous'' hypothesis --- our users appear to 
continually
seek out new and different
communities, and adopt the language characteristics of the new communities.  Another interesting point, albeit arguably less surprising,  is that  they tend to move to smaller communities (a fact
noted by Redditors\footnote{One comment: ``the longer you are on reddit, the more you get pulled into smaller subs''.}), which
might be a signal to site designers to make sure to offer a menu of narrowly-targeted
options for users to choose from (or to ensure that sub-groups can arise organically).
Finally, a complete surprise is that
for users who made at least 50 posts,
the patterns exhibited by those who
end up departing the site altogether are \emph{already} significantly different from those users who
end up
staying by {\em their first 10 posts}.  The fact that future abandonment can be
detected so early should be of interest to administrators of social-media
systems.  But, there is an unexpected factor potentially making this
discrimination difficult: in our data, the eventually departing users
are often most similar {\em not} to the least active users
in our study,
but to the {\em most} active users. We conjecture that
our ``dying'' users
are actively striving to remain engaged, but are not quite managing to explore
enough to make their overall posting experience satisfactory.  A design implication
might be to include mechanisms in one's site that more proactively suggest new,
diverse  sub-communities
for posting.

In Section \ref{sec:prediction},
we show that the
aforementioned differences in patterns are not ``mere'' correlations, but
do indeed serve as features that are effective at predicting future
activity level.

Again, our overall goal is to encourage further work on multi-community settings.  As a spur to the
imagination, and as a demonstration that this research domain is rich with
possibilities, we discuss in
sections   \ref{sec:singlemulti} and \ref{sec:langdiff} two additional questions
that arise.
First, what makes a user abandon a community and move on to new ones?
We see that the positivity of initial feedback
correlates with what groups users choose to return to, a finding that
contradicts recent results on the power of negative feedback \cite{cheng2014community},
albeit for 
commenting instead of posting.
Second, we make a foray into the ``situation vs. personality'' debate in
psychology \cite{Kenrick:AmericanPsychologist:1988,person-situation-09}: how much of our behavior is determined
by fixed personality traits, versus how much is variable and influenced by the
specific situation at hand? We consider this question from a linguistic
perspective, and determine that {\em even after topic-specific vocabulary is discarded} (after all, it wouldn’t be interesting to find that people use gym-related words at the gym that they don’t use at work),
users \emph{do} employ different language patterns in different communities.
This means that they are able to adapt even into ``maturity'' --- a positive
note to end on.

\section{Experimental setup}
\label{sec:data}

In the following, we first
describe
the data
that we use
and then propose
an analysis framework for capturing
the temporal dynamics
of multi-community engagement.

\subsection{Datasets.}

The main dataset used in this paper is
drawn
from Reddit,
a
very active
community-driven platform for submitting, commenting
on, and rating
posts\footnote{A Reddit post  consists at a minimum of a title that serves as anchor-text for a link.  The link may be to an offsite item (``link post'') or to some text that the post's author places on Reddit (``text post'').
The dataset with more detailed explanation is available at \url{https://chenhaot.com/pages/multi-community.html}.}
\cite{Singer:Www14:2014}.
Reddit is organized into thousands
of topic-based, user-created discussion forums called ``subreddits'',
which
users can post to essentially at will (modulo spam filtering, rate limits, and deletion of posts by moderators).
Other users can ``upvote'' or ``downvote'' posts; the difference between the number of upvotes and the number of downvotes, a difference that we henceforth refer to as {\em feedback}, is readily available.\footnote{The actual number of upvotes or downvotes is purposely inaccessible: \url{http://bit.ly/1xrciQY}.
}

Relying primarily on RedditAnalytics\footnote{\url{http://redditanalytics.com/}}, 
in February 2014 we collected all\footnote{
Except that we filter 
out
bots and banned users.
}
76.6M
posts ever submitted to Reddit since its inception,
together with their associated feedback values.  We discarded the last month of posts, since  their feedback values might not have had sufficient time to converge.
Since we need our users' community trajectories to be long enough to be
\emph{able}
to exhibit significant wandering (whether or not they actually do), the set of users we consider are those who have made at least 50 posts, following the choice in \cite{Danescu-Niculescu-Mizil:2013:NCO:2488388.2488416}.
We focus on the 157K \emph{\major posters}
who
first posted
between
 January 2008 and January 2012
so that
we have at least two years' worth
(2012-2014)
of observations for each of them.
We chose to start from January 2008 because
users were granted the ability to
create their own subreddits
at will
then.
Not only are the \major posters good objects of study because
we have a lot of data on their behavior,
but they also play a major role in determining the character of Reddit because they made
63\%
of the posts
written by users who first posted in the time period under consideration.%
\footnote{Cross-posting (posting the same
URL to multiple subreddits,
with or without a title change)
accounts for only 3\% of the posts from
the
users that we consider in this paper ---
only 1.77\% if we only consider their first 50 posts.
}
In order to ensure that our findings generalize beyond Reddit, we also consider a (more) physical-world
multi-community situation: the set of conferences in computer science.
Conferences generally correspond to topic areas within CS, and each can be
thought of as representing a social group, at least to some degree.
In this setting, we take ``posting'' to mean publishing a paper. We use the DBLP
database\footnote{%
\url{http://dblp.uni-trier.de}}
to find what papers appeared in which conferences, and refer to the resultant dataset as ``DBLP''.
For DBLP, we do not consider an analog of Reddit's feedback, although citation or
download counts could be used in future work.

It is important to note that
program committees play a huge role in determining an author's conference
trajectory.  This property makes
DBLP a less suitable domain for the
questions of user choice that we focus on
in this paper.
We thus place
our
DBLP trajectory results in the Appendix (\S \ref{sec:appendix}).

Statistics on the \major posters in Reddit and DBLP are given in Table \ref{tb:statistics}.
\newcounter{note}\setcounter{note}{1}
\newcounter{postnote}\setcounter{postnote}{\value{note}}
\para{Note \arabic{note}: how we define ``posting''.}\addtocounter{note}{1}
In this paper, we use the term {\em posting}
to refer to submitting an item to be voted or commented upon.  We distinguish
posting from {\em commenting on posts}
for several reasons.
First, posting is important for site designers to encourage since the site will presumably die without fresh conversation-starters.
Second, posting is not affected by a confounding factor that commenting is subject to:
Reddit
influences
commenting by how it presents potential targets for
comments (e.g., by
ranking them, or featuring targets on the Reddit home page).
Third, the way that comments are presented on Reddit makes scraping the complete commenting history rather difficult.
Nonetheless,
looking at commenting in multi-community environments is
an interesting
direction
for future research.
We conjecture that it would lead to new findings since, for example, we do know that top posters are generally not top commenters, and vice versa.%
\footnote{\url{http://bit.ly/1tendtD}
}

\newcommand{\amed}{.\textcolor{white}{00}} %
\begin{table}[t]%
\centering
\begin{tabular}{lrr}
& Reddit & DBLP\\ \hline
Average number of posts & 152.04 & 86.30\\
Median   & 89\amed & 71\amed \\ \hline %
Avg. no. of communities  & 28.85 & 38.08\\
Median  & 26\amed & 34\amed\\ \hline
Mean avg. time gap  btwn posts & 10.47 days & 3.36 mos\\ %
\hline
\end{tabular}
\caption{Statistics for \major posters (%
157K in Reddit,
10K in DBLP).
\label{tb:statistics}}
\end{table}

\subsection{Analysis framework.}

We now set up terminology and concepts that facilitate  discussion of users'
{trajectories} among communities.

{\em For each post by a \underline{given} user}, we store the
timestamp, \emph{time}, and the \emph{community}
(sometimes $C$ for short).  For Reddit data, we also store the post's
{\em feedback}
as of February 2014 and its \emph{\words} (the anchor-text plus any text written
by the user, all tokenized and part-of-speech tagged using the Stanford NLP
package\footnote{\url{http://nlp.stanford.edu/software/corenlp.shtml}}).

\begin{figure}[t]
\centering
\resizebox{0.48\textwidth}{!}{%
	\begin{tikzpicture}[font=\large]
	\draw (0,0) -- (15.01, 0);
	\draw[snake=ticks,segment length=1cm] (0,0) -- (15.1,0);
	\node[below] at (0,0) {$0$};
	\node[below] at (5,0) {$50$};
	\node[below] at (10,0) {$100$};
	\node[below] at (15,0) {$150$};
	\foreach \tick in {0,...,14}
	  \pgfmathtruncatemacro{\tickplusone}{\tick + 1}
		\draw [decorate,decoration={brace}] (\tick,0.2)  -- (\tickplusone,0.2)
   		node[midway,above] { $W_{\tickplusone}$};
  \foreach \tick in {0,...,4}
	  \pgfmathtruncatemacro{\tickplusone}{\tick+1}
	  \pgfmathtruncatemacro{\tickstart}{\tick*3}
	  \pgfmathtruncatemacro{\tickend}{\tick*3+3}
		\draw [decorate,decoration={brace}] (\tickstart,1.0)  -- (\tickend,1.0)
   		node[midway,above] { $S_{\tickplusone}$:20\%};
  \end{tikzpicture}%
}%
\caption{
Illustration of windows and stages for window size $w=10$, number of stages $S=5$, number of posts $T=150$, number of windows $\numWindows=15$. $W_i$ is a window; $S_i$ is a stage.
\label{fig:visualwindow}}
\end{figure}
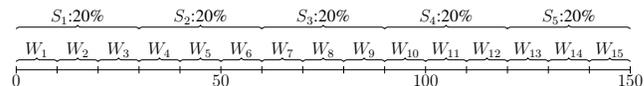

Several of the questions we are interested in pertain to properties of subsequences
of trajectories.  For example, suppose we want to know whether users are visiting
a broader set of communities over time; one way to check is to look at how
many communities they engaged with in their first $w$ posts
versus in their last $w$ posts.
Therefore,
a basic element in our analysis is a \emph{window}.  Let
variable $t$
index the posts made
by a user $u$, and suppose $u$ has made $T$ posts altogether.
We split the entire index
sequence
$1, \ldots, T$ into
non-overlapping
consecutive
 windows
$W_i$
 of size $w$,
where $i$ ranges from 1 to
$\numWindows\stackrel{def}{=}\floornum{T/w}$.
For example%
,
 in \shortfigref{fig:visualwindow}, $W_6$ would be the integers in
the
range $[51, 60]$.
We use $w=10$ throughout this paper.
Our Reddit results were insensitive to choices of $w$.

We define functions
$F$ on windows $W_i$ to summarize
properties
of that window and track how these
properties
change over time.
We use two ways to define $F$.
One way is to
directly
define
$F$ based on the entire window,
for example,
$F(W_i)=|\{
C_t: t \in W_i\}|$, the number of unique communities in $W_i$.
The other way is to define a function $f$ for each index $t$
--- for example, $f(t)$ could be the number of words in the $t^{th}$ post ---
and let
$F(W_i)$
be induced by $f$'s average value over the indices in $W_i$,
$F(W_i) = \frac{1}{w}\sum_{t \in W_i} f(t)$.
\begin{figure*}[t]
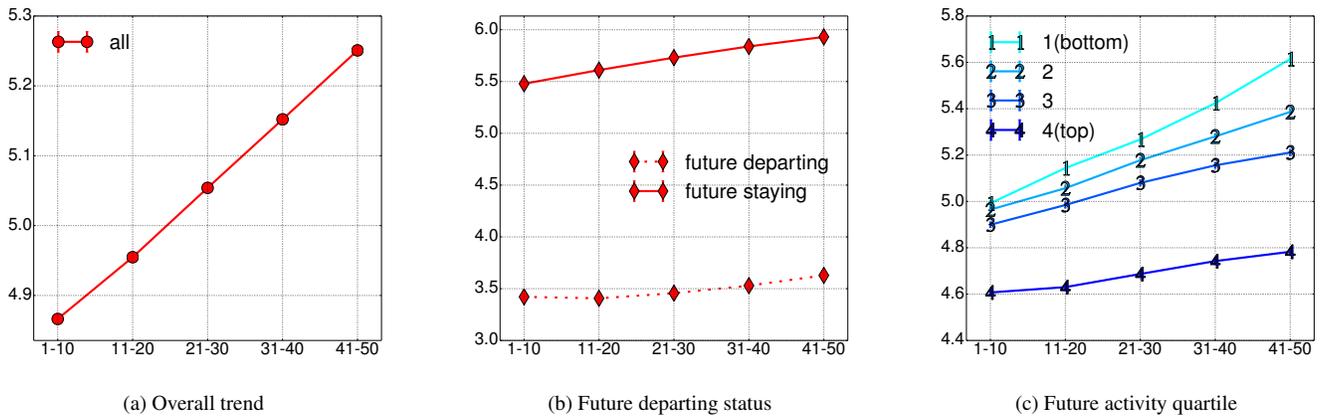

\centering
\subfigureplot{{plots/1107/sub_count/absolute_all_50_10_10_5}.pdf}{Overall trend}{fig:sub_count_all}{0.30}
\hfill
\subfigureplot{{plots/1107/sub_count/absolute_dl_50_10_10_5}.pdf}{Future departing status}{fig:sub_count_dl}{0.30}
\hfill
\subfigureplot{{plots/1107/sub_count/absolute_cmp_50_10_10_5}.pdf}{Future activity \level}{fig:sub_count_activity}{0.30}
\caption{Number of unique communities
per window.
x-axis: each of the first 5 windows.
y-axis:
number of unique communities appearing in the corresponding window.
In \shortfigref{fig:sub_count_dl} and \shortfigref{fig:sub_count_activity}, users are categorized
by their future state \emph{after} the initial 50 posts.
Standard-error intervals are depicted, but very small.
\label{fig:sub_count}}
\begin{fminipage}{\textwidth}
\vspace{-1.5mm}
\para{Note \arabic{note}:
y-axes scales%
, and other considerations regarding subsequent figures.
} \addtocounter{note}{1}%
Since many of the figures in 
Section \ref{sec:exploration} tend to support the same overall point
as in Figure \ref{fig:sub_count}%
, we make the subsequent figures relatively small 
({%
labeling the y-axes in the captions}),
but use the same x-axis, legends, and line styles in all of them.

As in Figure \ref{fig:sub_count},
each of the other figures in 
Section \ref{sec:exploration}
consists of three sub-figures.
In each, we scale
the
y-axes according to
the corresponding
data's range in order to show
significant changes
(all figures show standard-error bars, which are tiny).
But it should be noted that
the
lines when averaging over all users (leftmost sub-figure in the figures) would usually look flatter if plotted on the graphs that divide users by departure status (middle sub-figures) or activity \level (rightmost sub-figures).
\end{fminipage}
\end{figure*}

Given a window size $w$ and a function of interest, $F$, we take two perspectives to track the trajectory of $F$:
a
\emph{\relative view (all
the user's posts)}
and
a
\emph{%
\absolute view (%
50 posts)}.
The rationals are as follows:

The first
perspective%
, \emph{\relative},
tracks users' entire lifetimes.
Because the value of some functions is affected by choice of window size
 (e.g., the number of unique communities), we
still
fix
the
window size
in the \relative view,
but set an additional parameter $S$
of
the number of
life stages that we want to examine,
where each life stage contains the same number of windows, as depicted in \shortfigref{fig:visualwindow}.
For each stage, we compute the average
value over the windows in that stage.
A slight problem with
the
\relative view
is that
for different users,
the value of
the same life stage (say, the first 10\% of one's life)
may be based on a significantly different number of posts (say, 
10
for one user but 
100
for another).
The \relative view
also includes information
about the entirety of the user's life,
and thus is not appropriate for prediction settings
(for example, one does not ordinarily know at the time what percent of one's life has already passed).
Thus we also take a
\emph{\absolute} view, where
\emph{only}
the initial $50$ posts are examined.
(Recall from
the caption of
\shortfigref{fig:cumu_sub_count} that this encompasses a long time span on average.)
Thus,
the same amount of data is used for every user and
the induced features are valid for predicting future behavior.
For space reasons, in the main paper we will focus on the \absolute view, and
place some
\relative-view
results
in the Appendix (\S \ref{sec:appendix}).

\para{Future activity level.}
We
further
relate our analysis to users' future
activity level,
since future activity level is a useful quantity to predict.
We employ two different ways to categorize users' future commitment:
the two-way
classification of whether a user eventually abandons the global community altogether or not,
and
a 4-way split based on the
relative
number of posts that a user
eventually makes over his/her lifetime,
as follows.

\begin{itemize}
\item Departing status.
To determine
which users should be considered to have abandoned the site, we define
a date (specifically, 6 months before January 2014)
as
the
start-of-future
(SOF).
We define
\emph{departing}
users as those who stopped posting
as of SOF;
we define \emph{\staying} users as non-departing users who additionally post
at least once in the first 3 months and at least once in the second 3 months since
SOF, so that they are consistently ``active''.
There are 43,910 departing users and 75,708 \staying users.
Note that they all made at least 50 posts before
SOF.
\item Activity \level.
We split users into four quartiles based on the number of posts that they make in their entire life
 after the initial 50 posts.
(As it happens, the \staying/departing ratio is higher in the \higher \level.)
\end{itemize}

\section{%
Trajectory properties%
}
\label{sec:exploration}

We have established in \shortfigref{fig:cumu_sub_count} that users do
constantly
``wander around'' in multi-community environments.
In this section, we apply the framework proposed in \S \ref{sec:data} to explore three aspects of
this wandering process:
(\S\ref{sec:community}) the communities
users post to%
; (\S\ref{sec:languse}) the language users employ in
each community;
(\S\ref{sec:feedback}) the
feedback that users receive from other community members.
In \S \ref{sec:prediction},
we will further validate the effectiveness of
features
based on these properties
in prediction tasks.

\subsection{Multi-community aspects%
}
\label{sec:community}

We have shown in \S \ref{sec:intro} that users
on average consistently
post to 2.5
new communities every 10 posts (\shortfigref{fig:cumu_sub_count}).
But what else characterizes their patterns of movement among communities?  The answers to this  question have the design implications outlined in \S \ref{sec:intro}.

\para{Section summary.} \emph{We find that over time, users span more communities every 10 posts,
``jump'' more, and concentrate less.\footnote{
The continual exploration is not simply an effect of
the introduction of new communities over time.
For instance, although new communities or options also emerge in real life, people seem to settle down and do not explore much.
}
They enter smaller and less similar communities.
Eventually-%
departing users seem consistently less ``adventurous'' than \staying users even,
notably, from the very beginning.
Curiously, eventually-departing users act similarly to users in \high activity \level.
}

In the following, we explain the metrics for understanding these properties and discuss related theories.

\begin{figure}[h]
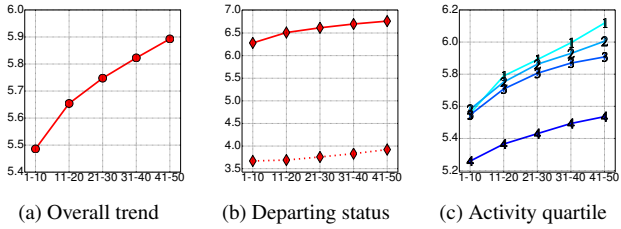

\centering
\subfigureplot{{plots/1107/sub_switch/absolute_all_50_10_10_5}.pdf}{Overall trend}{fig:sub_switch_all}{0.15}
\hfill
\subfigureplot{{plots/1107/sub_switch/absolute_dl_50_10_10_5}.pdf}{Departing status}{fig:sub_switch_dl}{0.15}
\hfill
\subfigureplot{{plots/1107/sub_switch/absolute_cmp_50_10_10_5}.pdf}{Activity \level}{fig:sub_switch_activity}{0.15}
\caption{Number of ``jumps''.
\label{fig:sub_switch}}
\end{figure}

\para{Users span more and more unique communities in a window, but
relatively speaking,
departing users span fewer unique communities.}
Figure \ref{fig:sub_count} shows the
per-window
number of unique communities that users post to.
The actual number is
interesting:
in \shortfigref{fig:cumu_sub_count}, users post to 2.5 new communities every 10 posts; here on average, users post to around 5 communities every 10 posts,
and thus only around 2.5 of them are ones that they have
ever
posted to.
Given that users have more potential communities to go back to over time,
this suggests that they do not tend to return to
some previous
communities.
More discussion
as to why
users
return to certain communities will be presented in \S \ref{sec:singlemulti}.
\para{%
Users ``jump''
between
communities more and more ``frequently'', but departing users do so
at around half the ``rate''.
}
(\shortfigref{fig:sub_switch})
To understand
how often users ``jump'',
we count the number of
``jumps'' that users make 
per window.
Formally, define $F(W_i)=\sum_{t, t+1 \in W_i} I(C_t \neq C_{t+1})$,
where
$I(x)$ is the indicator function:
$I(x)=1$ if $x$ is true, 0 otherwise.

Note that the number of unique communities in a window of 10 does not determine how often users
``jump''.
Given a window size of 10,
users can
jump as many as 9 times;
given that users on average span 5 communities in a window,
users can jump as few as 4 times.
In fact, users make around 5.8
``jumps'' per 10 posts.

\begin{figure}[h]
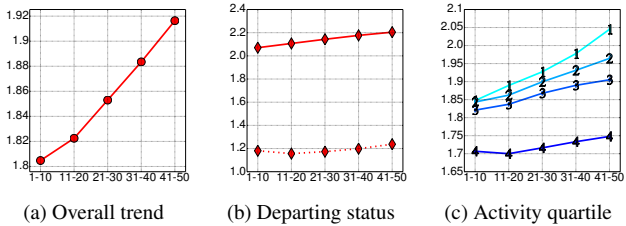

\centering
\subfigureplot{{plots/1107/sub_entropy/absolute_all_50_10_10_5}.pdf}{Overall trend}{fig:sub_entropy_all}{0.15}
\hfill
\subfigureplot{{plots/1107/sub_entropy/absolute_dl_50_10_10_5}.pdf}{Departing status}{fig:sub_entropy_dl}{0.15}
\hfill
\subfigureplot{{plots/1107/sub_entropy/absolute_cmp_50_10_10_5}.pdf}{Activity \level}{fig:sub_entropy_activity}{0.15}
\caption{Entropy of community-posting distribution.\label{fig:sub_entropy}}
\end{figure}

\para{%
Users spread their posts out more and more evenly, but relatively speaking, departing users focus more.}
(\shortfigref{fig:sub_entropy}%
)
We employ entropy as a metric for concentration%
,
following \cite{Adamic:2008:KSY:1367497.1367587}.
Entropy is based on the probability of a community appearing in a window $W_i$, $p_c = \frac{1}{w} \sum_{t \in W_i} I(C_t=c)$, and is defined as
$-\sum_c p_c \log_2{p_c}$ for $W_i$.
It is an information-theoretic measure that grows as the intra-window community-posting distribution approaches the uniform distribution (minimum concentration) \cite{Shannon:1948}.
The same qualitative
results hold if we use
the
Gini-Simpson index ($1-\sum_{c} p_c^2$), a commonly used metric in ecology for species concentration \cite{gini1912variabilita,simpson1949measurement}.%
\footnote{%
An alternative
hypothesis regarding the difference in activity \levels
is that
there isn't really a difference, but perhaps
 users in \higher \level
make
several posts in a single community where a lower-activity user makes just one,
e.g., $C_1C_1C_1C_2C_2C_2$ vs. $C_1C_2$.
If this were so, we would
observe a lower entropy simply due to
accidentally choosing a window size that is small relative to the average burst size.
However,
we verified that
this ``burstiness'' hypothesis does not hold, since
\higher users only
change communities %
about 0.5 fewer times than lower-activity
ones.
}

\begin{figure}[h]
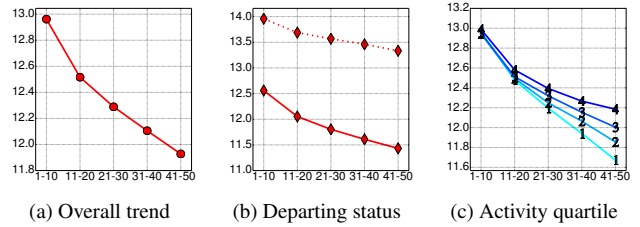

\subfigureplot{{plots/1107/sub_size/mean_log_size_absolute_all_50_10_10_5}.pdf}{Overall trend}{fig:sub_mean_size_all}{0.15}
\hfill
\subfigureplot{{plots/1107/sub_size/mean_log_size_absolute_dl_50_10_10_5}.pdf}{Departing status}{fig:sub_mean_size_dl}{0.15}
\hfill
\subfigureplot{{plots/1107/sub_size/mean_log_size_absolute_cmp_50_10_10_5}.pdf}{Activity \level}{fig:sub_mean_size_activity}{0.15}
\caption{Average $\log_2$(number of monthly posts in communities that a user posts to).
Note that it is {\em not} the case that big subreddits are being abandoned as a whole:  despite the availability over time of more and more small subreddits, the number of posts in the popular subreddits continues to increase.
\label{fig:sub_mean_size}}
\end{figure}

\para{Users enter smaller-looking communities (fewer posts per month), but
relatively speaking,
 departing users prefer larger communities.} (\shortfigref{fig:sub_mean_size})
Engaging with different communities entails
a choice between communities of different sizes.
A large community
can encompass diverse
community purposes and member preferences,
leading to broader appeal,
but at the same time,
a large size may dilute personal connection and lead to more conflicts \cite{Ren:07}.
Or, size might not have any effect at all.
To study this question, we set $f(t)$ to
log of the
number of posts made
by the user in the community in month $t$
as a simple metric of
how ``large'' the active portion of a community looks to an incoming user. %
\footnote{%
Reddit does not provide directly applicable metrics: the number of subscribers or those ``online now'' can consist mostly of passive observers.
The number of users who posted in a month is not presented at all, but we observe similar trends when extracting that as the metric.
}

We note that with respect to this metric of community size, the \relative view,  shown in the Appendix (\shortfigref{fig:sub_rel_mean_size_activity}), differs from the \absolute perspective plotted
above. In the \relative view,
\higher \level users
eventually
enter smaller communities
than lower-activity quartile users.
It seems that
they
just move more slowly to
such communities.

\begin{figure}[h]
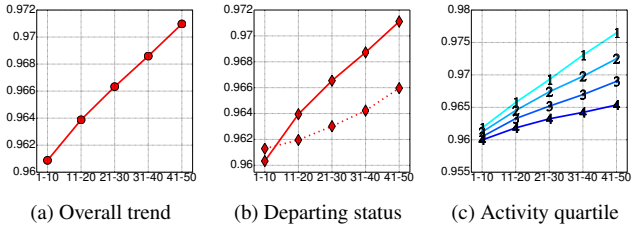

\subfigureplot{{plots/1107/sub_simi/user_absolute_all_50_10_10_5}.pdf}{Overall trend}{fig:sub_user_simi_all}{0.15}
\hfill
\subfigureplot{{plots/1107/sub_simi/user_absolute_dl_50_10_10_5}.pdf}{Departing status}{fig:sub_user_simi_dl}{0.15}
\hfill
\subfigureplot{{plots/1107/sub_simi/user_absolute_cmp_50_10_10_5}.pdf}{Activity \level}{fig:sub_user_simi_activity}{0.15}
\caption{Community dissimilarity based on
poster
overlap.\label{fig:sub_user_simi}}
\end{figure}

\para{Users post to less similar communities over time, but relatively speaking, departing users prefer more similar ones.}
(\shortfigref{fig:sub_user_simi}%
)
One hypothesis for
how people
select new communities is that
they explore
similar communities
to those they have visited in the past,
because they want more exposure to topics that they are already interested in.
On the other hand,
perhaps they choose new communities because their
interests have changed,
implying that they would choose more different communities.

We measure the
dissimilarity 
between communities
$C_1$ and $C_2$
based on poster overlap,
restricting attention to just those communities with at least 1000 posts to ensure sufficient data.
Denoting the set of users who ever posted in a community
$C$
 as $U_C$,
our measure is
$1-\frac{|U_{C_1} \cap U_{C_2}|}{|U_{C_1} \cup U_{C_2}|}.$
Note that the dissimilarity between two communities
is
computed based on their eventual poster set,
since we want to capture the ``actual'', eventual relationship between the two,
and
so
does not change over time.
For a window $W_i$, the overall community
dissimilarity $F(W_i)$ is defined as the average of all the pairwise
dissimilarities between
the
communities
that the user posted at during that window $W_i$.

The same trends
hold if we measure language
dissimilarity between
communities using
the
KL-divergence between community language models.

\para{Different activity \levels.}
For \emph{all} of the above metrics, users
of different {\em future} activity \levels manifest significant differences
even in their very earliest behavior,
although the
differences are not as dramatic as those between departing users and \staying users.
The curves for the different \levels always appear in either the order
1,2,3,4 or 4,3,2,1,
and the highest-activity \level
curves are always the closest to those for departing users.

\begin{figure}[t]
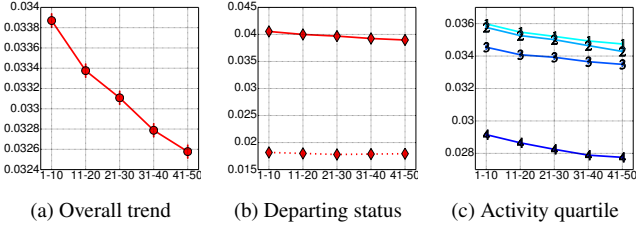

\subfigureplot{{plots/1107/sub_words/i_absolute_all_50_10_10_5}.pdf}{Overall trend}{fig:sub_i_all}{0.15}
\hfill
\subfigureplot{{plots/1107/sub_words/i_absolute_dl_50_10_10_5}.pdf}{Departing status}{fig:sub_i_dl}{0.15}
\hfill
\subfigureplot{{plots/1107/sub_words/i_absolute_cmp_50_10_10_5}.pdf}{Activity \level}{fig:sub_i_activity}{0.15}
\caption{Percentage of first singular person pronouns.\label{fig:sub_i}}
\end{figure}

\subsection{Language aspects}
\label{sec:languse}

The second aspect that we examine is the language that users employ
within communities.
This
examination, and the formulation we apply below,
are inspired by
\cite{Danescu-Niculescu-Mizil:2013:NCO:2488388.2488416},
which found that in
single-community settings, users first pass through
an ``adolescent'' phase where they learn linguistic norms,
but after this phase
stop adapting to new norms and become
increasingly
distant from
the community.
Our results
indicate
that this is \emph{not} the case in
the
multi-community setting.
Rather, with respect to part-of-speech tags or stopwords, users do not move farther and farther away from
the community distribution; and
when (frequent) content words are included,
users seem to ``stay young'',
continuously growing closer to the
community's language.
Surprisingly, departing users are better mimics of the community's language than 
\staying 
users
are.
The bulk of this section provides the experimental evidence, based on various
forms of cross-entropy,  from which we draw
these conclusions.

Additionally, we, like \cite{Danescu-Niculescu-Mizil:2013:NCO:2488388.2488416},
find that the usage of 1st-person-singular pronouns (e.g., I, me) declines over
time,%
\footnote{Acronyms such as ``TIL'' (for ``today I learned'') were not included.}
which has been argued to indicate a greater sense of community
affiliation \cite{chung2007psychological,Sherblom:CommunicationResearchReports:2009}.  However, upon closer inspection, the fact that departing posters use these
words {\em less} frequently than those users who end up staying seems
problematic for such theories --- although one could speculate that the cause is
that our departing users start out with strong affiliation needs but become
disappointed.  These results are shown in \figref{fig:sub_i}.

\para{Cross-entropy with vocabulary-varying language models}.
We use cross-entropy to measure the distance between
(a language model constructed from) a user's $t^{th}$ post and a
language model built from  all the posts in the corresponding community, $C$, in that
same month $m(t)$.  Importantly, we will compute these models based on various
choices of vocabulary $V$; this will reveal that although users' topical-word
usage grows closer and closer to that of the community's, their
usage in part-of-speech tags and stopwords stabilizes in terms of distance from the community's.

The first step of our $V$-dependent language-model construction is to replace
every instance of any word not in $V$ with the new token ``\unk''.  Next, we
define the
community-based
language model to be the distribution  over $v \in V \cup \{\unk\}$ given by setting
$p^{C}$
to the relative frequency of $v$ in the concatenation $\words^{C, m(t)}$ of all the posts in $C$
during the month $m(t)$.  Then, we measure the cross-entropy by
$$f(t)=
\frac{1}{|\words_t|}\sum_{v \in \words_t} \log_2 \frac{1}{p^{C}(v)}.$$
(This equation shows why we do not need to smooth the community language model:
since $\words_t$ is a component of $\words^{C, m(t)}$, $p^{C}(v) > 0$ for $v \in
\words_t$.)

With all of this in hand, \figref{fig:sub_ce} depicts representative evidence for the conclusions we drew at the
beginning of this section.  Specifically,  the evidence consists of cross-entropy values for $V$ chosen to be 46
parts-of-speech tags, the most frequent 100
words in
Reddit, or the most frequent 1000 words in Reddit.
 Trends for
$V$ set to the 500 or 5000
most frequent words are similar to the most frequent 1000 words.

\newcommand{\manline}{\smallskip\line(1,0){220}\smallskip}

\begin{figure}[th]
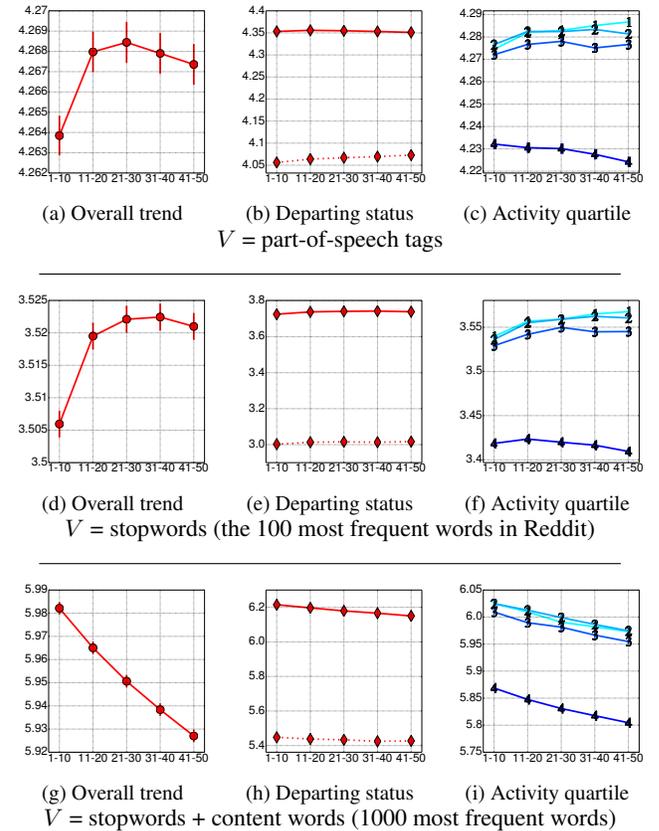

\centering

\subfigureplot{{plots/1107/sub_ce/pos_ce_absolute_all_50_10_10_5}.pdf}{Overall trend}{fig:sub_pos_ce_all}{0.15}
\hfill
\subfigureplot{{plots/1107/sub_ce/pos_ce_absolute_dl_50_10_10_5}.pdf}{Departing status}{fig:sub_pos_ce_dl}{0.15}
\hfill
\subfigureplot{{plots/1107/sub_ce/pos_ce_absolute_cmp_50_10_10_5}.pdf}{Activity \level}{fig:sub_pos_ce_activity}{0.15}

$V$ = part-of-speech tags
\manline

\subfigureplot{{plots/1107/sub_ce/100_ce_absolute_all_50_10_10_5}.pdf}{Overall trend}{fig:sub_100_ce_all}{0.15}
\hfill
\subfigureplot{{plots/1107/sub_ce/100_ce_absolute_dl_50_10_10_5}.pdf}{Departing status}{fig:sub_100_ce_dl}{0.15}
\hfill
\subfigureplot{{plots/1107/sub_ce/100_ce_absolute_cmp_50_10_10_5}.pdf}{Activity \level}{fig:sub_100_ce_activity}{0.15}
$V$ = stopwords (the 100 most frequent words in Reddit)
\manline

\subfigureplot{{plots/1107/sub_ce/1000_ce_absolute_all_50_10_10_5}.pdf}{Overall trend}{fig:sub_1000_ce_all}{0.15}
\hfill
\subfigureplot{{plots/1107/sub_ce/1000_ce_absolute_dl_50_10_10_5}.pdf}{Departing status}{fig:sub_1000_ce_dl}{0.15}
\hfill
\subfigureplot{{plots/1107/sub_ce/1000_ce_absolute_cmp_50_10_10_5}.pdf}{Activity \level}{fig:sub_1000_ce_activity}{0.15}

$V$ = stopwords + content words (1000 most frequent words)

\caption{Distance from the community language model.  The rows indicate
different choices of vocabulary $V$.
\label{fig:sub_ce}}
\end{figure}

\para{Technical aside: the potentially confounding factor of rare words
interacting with community posting
volume.}
We also
used a ``full'' vocabulary that contains all words that appear more than 100 times in Reddit (180K
types),
but do not show the results here.
This is due to the fact that for large vocabulary sizes, what
appears to be differences in language matching can actually be merely a side-effect of
one
class of users
posting in
more-voluble communities.
The
argument runs as follows.
The full vocabulary allows for many words $v'$ with low frequency in the
community --- say,
1 ---  to
contribute to the cross-entropy computation. The probability estimate $p^C(v')$ for such words is  $1/|\words^{C, m(t)}|$
(where $t$ is chosen appropriately).
So, in groups where $|\words^{C, m(t)}|$ is large, the contribution of such $v'$
to the cross entropy is bigger than it would be for sub-communities where $|\words^{C, m(t)}|$
is small.\footnote{
This concern cannot be alleviated  simply by sub-sampling a community's posts,
 since the true root of the problem is rare words, not
just the length and number of posts in the community per se.
}
\subsection{Feedback
aspects
}
\label{sec:feedback}

A final question
that Reddit data 
allow us to easily answer is,
how are users received by other members of the community?
For each post,
Reddit provides
the difference between
the number of
upvotes and
number of
downvotes.
Because
the average value of this difference can
vary
among different communities, we measure the feedback that users get by the relative position of this difference among all posts in the community that month, i.e., how often the posts made by a user 
outperform
the ``median post'' in a community.
For each index $t$, we define $f(t)$ as $I(feedback_t > median(C_t, m(t))),$ where $median(C, m)$ represents the median vote difference in community $C$ in month $m$.

\begin{figure}[h]
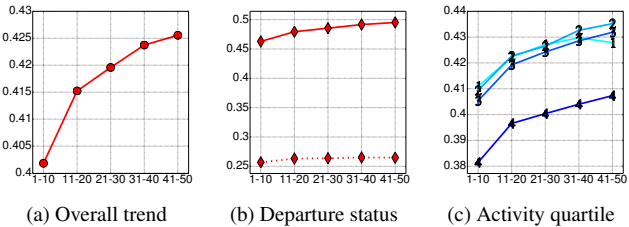

\subfigureplot{{plots/1107/sub_median/median_absolute_all_50_10_10_5}.pdf}{Overall trend}{fmediang:sub_median_all}{0.15}
\hfill
\subfigureplot{{plots/1107/sub_median/median_absolute_dl_50_10_10_5}.pdf}{Departure status}{fig:sub_median_dl}{0.15}
\hfill
\subfigureplot{{plots/1107/sub_median/median_absolute_cmp_50_10_10_5}.pdf}{Activity \level}{fig:sub_median_activity}{0.15}
\caption{%
Success rate at
outperforming the median vote difference.
\label{fig:sub_median}}
\end{figure}

Surprisingly, the feedback that \major posters receive is \emph{continually} growing
more positive,
although the rate slows
over time (\shortfigref{fig:sub_median}).
However, the growth is small compared to the drastic differences between departing users and 
\staying 
users.
Even departing users get
more-positive feedback over time, but
the increase is not as great as for 
\staying 
users.
Users in \high activity \level also fare worse, although as shown in the relative perspective (\shortfigref{fig:sub_rel_median_activity}), they
catch up in the later stages of their life.
The results are consistent if we measure how often posts outperform 75\% 
of the community's posts.

\subsection{Recap}
\label{sec:traj-summary}
In all three aspects that we examined, users with different future activity
levels manifest significant differences in their trajectories of multi-community engagement.
Interestingly, users that eventually depart seem ``destined'' to do so  even from the very beginning, since the curves for the departing vs. \staying users generally start out apart and maintain or increase that distance over time.
Meanwhile, there are smaller but significant differences in these metrics between users at different activity \levels.
It is
important to note that some metrics can be correlated (e.g., number of unique communities and entropy).
However, none of the metrics determines another,
so we believe
discussing each one of them
was valuable.

\begin{figure}[t]
\centering
\includegraphics[width=0.35\textwidth]{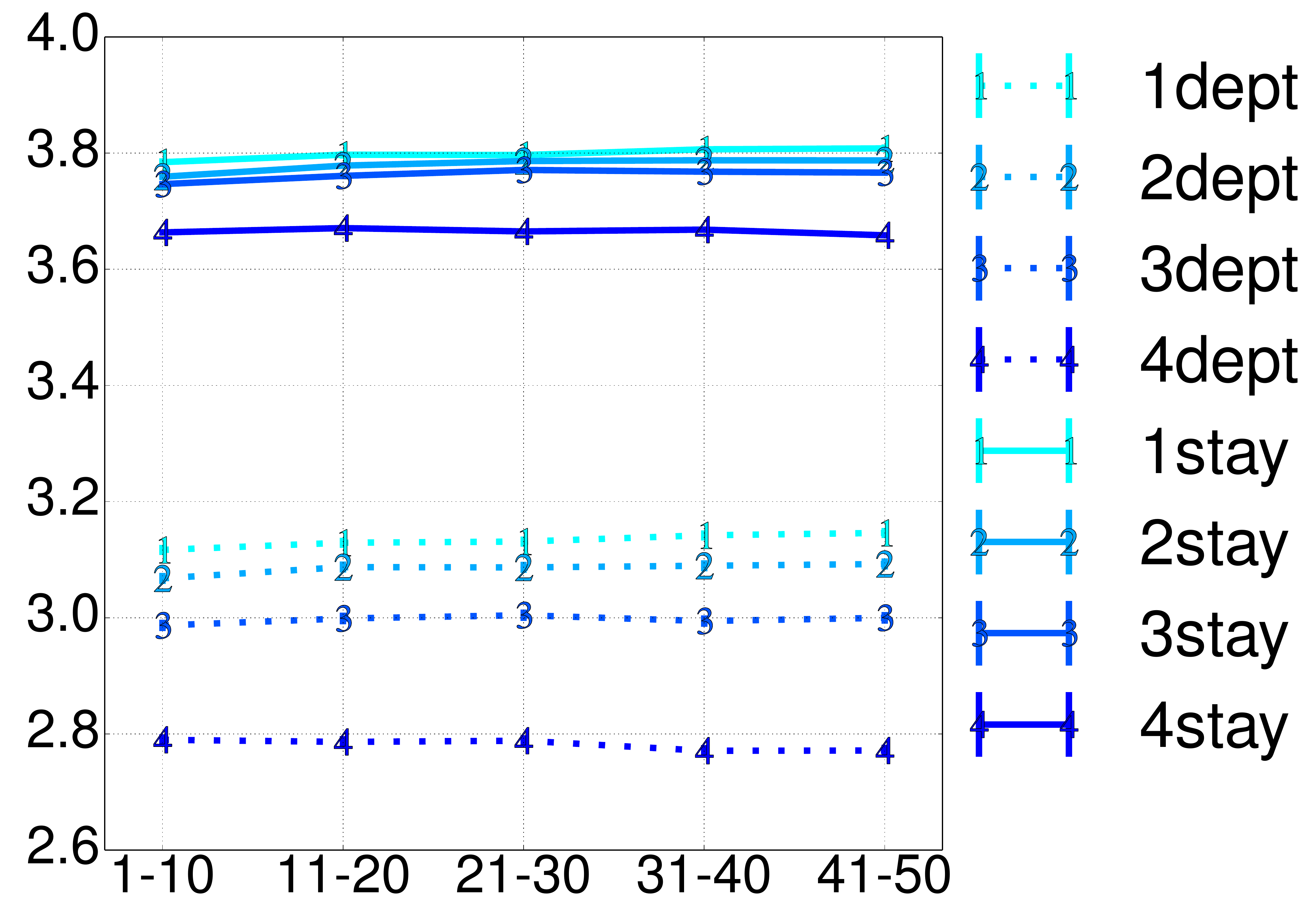}
\caption{Interplay between
departure status
and activity \levels.
y-axis:
distance from the corresponding monthly language model when setting the vocabulary to the
100 most frequent words. $i$dept refers to departing users in the $i$-th quartile; $i$stay refers to 
\staying 
users in the $i$-th quartile. \label{fig:interplay}}
\end{figure}

Another interesting phenomenon we consistently observe is that
for all our
metrics, users in \high activity \level are the closest to the departing users in the first 50 posts
(a direct comparison for language is shown in \shortfigref{fig:interplay}).

\newcommand{\fullpred}{1-prediction\xspace}
\newcommand{\winpred}{5-prediction\xspace}

\section{Predicting departure and activity levels}
\label{sec:prediction}

We have now seen many properties of multi-community engagement that correlate with user activity.
To examine the effectiveness of these properties in prediction,
we set up two different prediction tasks
that
correspond to how we measure users' future
activity level
in \S \ref{sec:data}:

\begin{itemize}
\item Future departure status.
In this task, we predict whether users abandon Reddit in the future.
We use F1 for evaluation, with the minority class (%
departing
users, as defined in \S \ref{sec:data})
 as the target class.
We use weighted L2-regularized
logistic regression as
classifier.
\item Future total number of posts. This is a regression task
where the goal is, for a given user, to estimate $\log_2$(future number of posts).
We employ L2-regularized support vector regression, and measure performance by
root mean squared error (RMSE).
\end{itemize}
Each instance consists of a user's first 50 posts.

\para{Baseline and features.}
We consider the following feature sets, where for window-based features we set the window size
$w=10$, thus deriving 50/10 = 5 values.\footnote{Alternatively,
one could
set $w=50$,
thus extracting features from all 50 posts in a single batch.
This approach
turns out to be
poorer than using 5 windows
because trend information is not captured.}

\begin{itemize}
\item
Average
time-gap between posts.
\cite{Danescu-Niculescu-Mizil:2013:NCO:2488388.2488416} states that
this is an effective feature used in prior work on churn prediction \cite{Dror:2012:CPN:2187980.2188207,yang2010activity}.
{\em Thus,  this feature by itself serves as our (strong) baseline.}
\item Multi-community
aspects
(%
henceforth ``sub info''). This includes number of unique communities, number of ``jumps'',
entropy, and Gini-Simpson index based on
the user's
community-posting distribution,
as well as
mean log
``apparent''
community size as defined in \S \ref{sec:community}.
Similarity between communities is not used
because information about the future is
incorporated
in
the way we compute it.
\item Language
aspects
(``lang'' for short). This includes cross-entropy with
the
monthly community language model
for the following choices of vocabulary:
part-of-speech tags; the top 100, 500, 1000, 5000, 10000
most frequent words;  and
the
full vocabulary as defined in \S \ref{sec:languse}.
Additionally, we include the proportion of 1st-person-singular pronouns and post length in words.
\item Feedback
aspects.
This includes the fraction of posts that outperform 50\% and 75\% of all
of the corresponding month's worth of the community's posts in terms of positivity of feedback. Refer back to \S \ref{sec:feedback} for more information.
\end{itemize}

For entropy, Gini-Simpson index, and number of unique communities, we %
include the value
for all 50 posts, since for these features, the values
for all 50 posts
are not simply the average of
the
values from 5 windows of 10 posts.
We also use the index of the window with the largest value and the smallest value as features,
following
\cite{Danescu-Niculescu-Mizil:2013:NCO:2488388.2488416}.
All features are linearly scaled to $[0,1]$ based on training data.

\para{Experiment protocol.} In both tasks,
we perform 30 randomized trials.  In each trial,
we
randomly draw
20,000 users from our dataset as training data and
a distinct set of 5,000 users as testing data. We use 5,000 users from
the
training data as validation
set.
We use LIBLINEAR \cite{Fan:2008:LLL:1390681.1442794} in all prediction tasks.
For significance testing, we employ the paired Wilcoxon signed rank test \cite{Wilcoxon:BiometricsBulletin:1945}.

The standard procedure for generating learning curves would be to only look at the {\em first} $x$ posts as $x$ varies,  $x=10,20,30,40,50$.
A non-obvious but ultimately fruitful idea we introduce here is to contrast the effectiveness of the information
in the early part of each 50-post instance with that of the late part of the 50-post instance.
That is,
we compare the performance
if we use the {\em first} (``fst'' in our plots) $x$ posts with the performance of using the {\em last} $x$ posts.
(One might expect later periods to be more predictive,
given that
they are more recent.
But surprisingly,
we will see that
when we predict departure status,
we find that earlier information is more useful, which again suggests that departing users are ``destined'' to leave from the
very
beginning.)

\subsection{Predicting departing status}
\label{sec:departure_prediction}

\para{%
Basic comparisons.} (Figure \ref{fig:departing_window})
Using all features outperforms
a strong baseline that uses time-gap features by 18.3\% --- the difference between an F1 of .699 and an F1 of .591 ---
which shows the effectiveness of
features
drawn from multi-community engagement.

The performance
of the first $x$ posts is always above that of the last $x$ posts.
This suggests that the initial information is more predictive
of eventual departure.
Note that for \major posters, departure is
quite
 ``far away'' from the initial posts.
In fact, using all features
drawn from only
the first 10 posts outperforms time-gap features
extracted from all
50 posts.
Thus it
may be
very important for
designers of social systems to make sure that users start well,
perhaps through positive feedback
or by recommending communities to post in
(which can differ from the communities one might recommend that a user reads).
\para{Feature-set analysis.} (Figure \ref{fig:departing_feature}) In predicting departure, it is most useful to know how well users
match
a community's language%
.
The second most useful
features are the patterns of community visitation.
Language-matching, community-trajectory, and community-feedback features
all outperform
time-gap information,
which suggests that how users interact with different communities is more predictive than activity rate in predicting whether
\major
users will leave.

\begin{figure}[t]
\centering
\subfigureplot{{plots/1107/prediction/death_1107_all}.pdf}{Window comparison}{fig:departing_window}{0.25}%
\subfigureplot{{plots/1107/prediction/death_1107_cmp}.pdf}{Feature comparison}{fig:departing_feature}{0.25}%
\caption{%
Results for predicting departing status.
y-axis: F1 measure.
In \shortfigref{fig:departing_window}, the dashed lines show the performance
of the baseline, timing-based features;
the solid lines show the performance of using all features.
Red lines show the performance using the first $x$ posts, while blue lines show the performance using the last $x$ posts.
\shortfigref{fig:departing_feature}%
:
performance of different feature sets.
All differences for 50 posts are statistically significant according to the Wilcoxon signed rank test ($p<0.001$).
\label{fig:departing}}
\end{figure}

\begin{figure}[t]
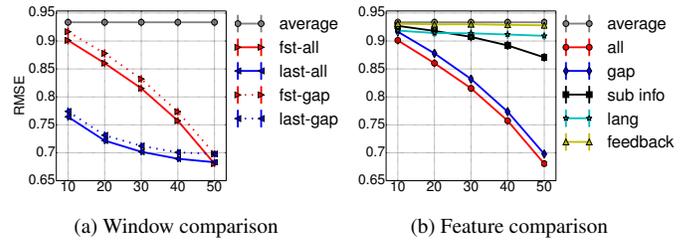

\centering
\subfigureplot{{plots/1107/prediction/activity_count_1107_all}.pdf}{Window comparison}{fig:activity_window}{0.25}%
\subfigureplot{{plots/1107/prediction/activity_count_1107_cmp}.pdf}{Feature comparison}{fig:activity_feature}{0.25}%
\caption{%
Results for predicting $\log_2$(future total number of posts). y-axis: RMSE, the smaller the better.
The line styles are the same as in \shortfigref{fig:departing}.
``Average'' shows a baseline that always predits the mean value in the training data.
All differences for 50 posts are statistically significant according to the Wilcoxon signed rank test ($p<0.001$).
\label{fig:activity}}
\end{figure}

\subsection{Predicting activity \level}

\para{Comparisons with the baseline.}
(Figure \ref{fig:activity_window},
\ref{fig:activity_feature})
In contrast to the case just discussed of
predicting departure status,
time-gap
between posts
is
a much stronger feature in predicting future total number of posts.
This is plausible because for these \major posters,
time-gaps in posting determine how many posts that people can physically make.
However, adding
all the
features based on multi-community engagement still improves the performance over
timing information to a statistically significant degree.
Prior work has
shown
that adding language features can lead to big improvements over timing-based features \cite{Danescu-Niculescu-Mizil:2013:NCO:2488388.2488416};
the relatively small improvement in our experiment
may be due to the fact that the
datasets in \cite{Danescu-Niculescu-Mizil:2013:NCO:2488388.2488416} have a longer history
than ours.

Also,
using the last $x$ posts is much more effective than using the first $x$ posts.
There thus seems to be different factors affecting
\major posters
with respect to deciding whether to remain in a community versus deciding to be highly active in it.
\section{When do users abandon their posts?}
\label{sec:singlemulti}

We have already seen that (our) users constantly try out new communities, but we have not yet addressed a related question
of practical importance to community maintainers, as well as of inherent social-scientific interest:
how much and why do users {\em abandon} communities?

We can frame the ``how much'' issue succinctly by asking the following
question.  Suppose we partition the set of communities a user visits into (1) those that
he or she abandons after just a single post, and (2) those that he or she posts
at least twice to. Which set --- the single-post communities or multiple-posts
communities, is larger, on average?
We claim that the answer is not a priori obvious\footnote{Recall the title of
Duncan Watts' recent book ``Everything Is Obvious: *Once You Know the Answer''%
.}.
But the data shows that users rack up more abandoned communities than return
engagements, as depicted in the figure below. This suggests that although users are
constantly willing to post to  new groups, they
are often only 
giving
 these new groups one shot.
\begin{figure}[h]
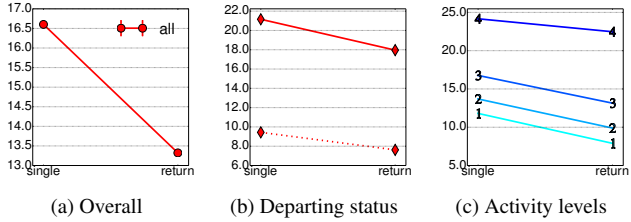

\subfigureplot{{plots/1107/single_multi/len_all}.pdf}{Overall}{fig:single_multi_len_all}{0.15}
\hfill
\subfigureplot{{plots/1107/single_multi/len_dl}.pdf}{Departing status}{fig:single_multi_len_dl}{0.15}
\hfill
\subfigureplot{{plots/1107/single_multi/len_cmp}.pdf}{Activity levels}{fig:single_multi_len_activity}{0.15}
\caption{
Comparison
of the average number of communities where a user posts only once
vs. more than once.
\label{fig:single_multi_len}
}
\end{figure}

What is happening in the single-post communities that causes a user to stop
posting in them immediately?  We find that positivity of feedback (in Reddit,
the difference in upvotes and downvotes) may play a substantial
role, as shown by the figure below.  \figref{fig:single_multi_mean} is based on the {\em very first}
post that a user makes in every community they posted in; it  plots the percentage of
such first posts that received a
feedback score above that of the median feedback score in the respective community.

The reason that this is interesting to note is
that our results contrast with previous findings of the power of {\em negative}
feedback for predicting repeated commenting \cite{cheng2014community}; we conjecture that the
difference is due to different impulses driving posting vs. commenting behavior.

\begin{figure}[h]
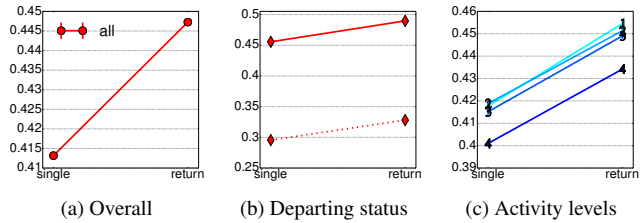

\subfigureplot{{plots/1107/single_multi/both_mean_all}.pdf}{Overall}{fig:single_multi_mean_all}{0.15}
\hfill
\subfigureplot{{plots/1107/single_multi/both_mean_dl}.pdf}{Departing status}{fig:single_multi_mean_dl}{0.15}
\hfill
\subfigureplot{{plots/1107/single_multi/both_mean_cmp}.pdf}{Activity levels}{fig:single_multi_mean_activity}{0.15}
\caption{Users get better feedback for the first post in the communities that
they eventually returned to than for the communities that they ended up making
only a single post in.
y-axis: average fraction of a user's post with feedback score better than the
community's median.
We exclude users that have only single-post communities or
only
multiple-posts communities,
thus controlling for individual-user characteristics to some extent.
All differences
between connected points are statistically significant
according to the
paired t-test ($p < 0.001$).
\label{fig:single_multi_mean}}
\end{figure}

\section{Do users speak differently in different communities?}
\label{sec:langdiff}

So far
we have
revealed interesting
and
sometimes
arguably counterintuitive
properties of multi-community engagement,
and demonstrated that they are effective cues in predicting a user's future activity level.
But an additional fascinating and orthogonal question is: when users participate in multiple communities,
to what degree are their actions stable \emph{across}
settings?
To look at this question is to contribute another piece of evidence to
the ``situation vs. personality'' debate \cite{Kenrick:AmericanPsychologist:1988,person-situation-09}: how much of our behavior is determined
by fixed personality traits, versus how much is variable and influenced by the
specific situation at hand? Or, to put it a  bit more dramatically, are you
fundamentally the same person at work as you are at the gym?

Here, we study the question with respect to language use.  The overall message is,
{\em even after topic-specific vocabulary is discarded} (after all, it wouldn't be
interesting to find that people use gym-related words at the gym that they don't
use at work), individuals  \emph{do} employ different language patterns in different
communities.  The way we determine this is conceptually straightforward:  we check whether it's possible to
tell which community a user's posts come from based just on the distribution of
stopwords or non-content-words within their posts.

 The rest of this section gives a quick sketch  of
our experiments. (Space constraints preclude a full discussion of the details.)

\medskip

If we fix some vocabulary $V$ of non-content words, then we can create
classification instances from the 227K triples that exist in our data consisting
of (1) a user $u$,
(2) words of $u$'s first 25 posts in some community $C_1$,
and (3) words of $u$'s first 25 posts in a different community $C_2$.
Then, we compute the cross entropy of each post
against
the corresponding monthly language models, over the restricted vocabulary $V$,
constructed from each of the two communities $C_1$ and $C_2$.\footnote{Actually, we divide these 25 posts into windows of 5 posts and take
the
average cross entropy in each window, in order to be more robust and potentially
capture trends, but it simplifies exposition to think of just
a single post.} Add-1/|V|
smoothing is applied to all language models concerned.
We then use these non-content-word cross-entropies as features to guess which of (2) and (3) came from community $C_1$.

We run experiments for several choices of $V$: parts-of-speech, the 100 most
frequent words in Reddit, and the 500 most frequent words in Reddit.  The first
two choices definitely do not include topic-specific words, and the latter will
not include many  (there are
180K words in the full Reddit
vocabulary), and so these choices may be taken to represent a user's language {
\em style} \cite{argamon2005measuring,DanescuNiculescuMizil:ProceedingsOfWww:2011}.
If the user's style does not change from community to community, then the
cross-entropy features mentioned above will not be helpful for determining that
item (1) comes from $C_1$ and not $C_2$; thus, accuracy at matching
language model to community would be 50\%. But, as shown below, the average accuracies, utilizing logistic classification,  of 30 random-split experiments (10K tuples for training and development, 2500 for
testing) for each choice of $V$
are (statistically) significantly above 50\%:
{\begin{center}
\begin{tabular}{lr}
$V$ & accuracy \\ \hline
parts of speech & 62.5\% \\
most frequent 100 words & 56.0\% \\
most frequent 500 words & 61.4\% %
\end{tabular}
\end{center}
}

\section{Related work}
\label{sec:related}

Anthropologists, psychologists and sociologists have looked at some questions regarding multi-community engagement,
often in the context of interaction with new social circles or cultures \cite{Buhler:JournalOfAppliedPsychology:1935,Hurtado:JournalOfSocialIssues:1997,Berry:AppliedPsychology:1997}.
Recently,
computer scientists
have turned to
examining
multi-community engagement data
available online \cite{Backstrom:2006:GFL:1150402.1150412,Adamic:2008:KSY:1367497.1367587,Vasilescu:IeeeXplore:2013,Vasilescu:2014:SQS:2531602.2531659,Lakkaraju+McAuley+Leskovec:13}.
Our work differs
by focusing
on the following specific
problems: (a) characterizing
full
community-trajectory
sequences,
as opposed to looking at pairwise community transitions \cite{Backstrom:2006:GFL:1150402.1150412,Vasilescu:IeeeXplore:2013,Vasilescu:2014:SQS:2531602.2531659};
(b) revealing how properties of these trajectories correlate with a user's future
cross-community activity
--- we incorporate but also go beyond language-based features, as inspired by previous within-community work \cite{Danescu-Niculescu-Mizil:2013:NCO:2488388.2488416,Rowe:Icdm:2013}, and timing-based features \cite{Dror:2012:CPN:2187980.2188207};
(c) considering the effect of each community's positive and negative feedback, which may shed light on why users choose some communities over others.

Researchers have
also
been working on predicting users' survival (also known as churn prediction) \cite{Dasgupta:2008:STR:1353343.1353424,Dror:2012:CPN:2187980.2188207,yang2010activity} and activity level \cite{DeChoudhury:2010:IRS:1772690.1772722,Zhu:2013:PUA:2505515.2505518}.
They focus on
the single-community setting.
A number of studies examined
community-level evolution or the success of
individual communities (often websites)
\cite{Iriberri:2009:LPO:1459352.1459356,Kairam:2012:LDO:2124295.2124374,Ludford:2004:TDI:985692.985772,Zhu:2014:SEN:2556288.2557348,Zhu:2014:IMO:2556288.2557213},
whereas our work focuses
on the life cycle of users.
\section{Concluding discussion}
\label{sec:conclusion}

\para{Summary.}
We have investigated properties of multi-community engagement;
this is a setting that has not received much computational research attention before, and yet is important because it
encompasses many online and physical situations.
In this first large-scale study of the phenomenon,
we have found a number of sometimes counterintuitive but robust properties --- some
involving choice of community, some involving language use within communities,
and some involving feedback from communities ---  revolving
around the discovery that users ``wander'' and explore communities to a greater
extent than might have been previously suspected.
\para{Limitations and further directions.}
We focused on
posting, but commenting and other
related behaviors
are very interesting subjects for future study.
Our study is quantitative and observational.
Qualitative studies, or
controlled experiments regarding the design implications in \S \ref{sec:intro}%
,
can further improve our 
understanding.
It is important to note that
our study is limited to
``\major
posters'' so that
we would have enough history
per
user to observe a relatively long
trajectory.
This is an unusually engaged group of users that comprises 5.9\% of our users.
We have not addressed the question of how  multi-community engagement is
exhibited by
users
who are not as active.
The notion of
considering users to exist in a multi-community setting can in principle be extended to looking at user behavior across multiple websites or apps.
With the advent and adoption of multiple-website services such as OpenID, observing users at that scale of multi-community engagement may well become quite important in the future.

There are many more challenging questions
that arise from taking a
multi-community perspective.
For example,
are the particularly nomadic treated differently?
What is multi-community engagement like in real life, considering the cost of switching?
How can we extend current theories and principles in community design to a multi-community setting?
Further
understanding of these questions is crucial for on- and off-line community design
and an exciting direction for future work.
\section{Appendix}
\label{sec:appendix}

\para{\Relative view for users in Reddit.}
In general, the overall trends and differences between departing users and staying users are the same
as in the \absolute view.
But in terms of activity \levels, there are some interesting differences.
For example,
the ordering
of the
activity \levels  with respect to mean
$\log_2$(number of posts that month)
completely reverses itself
(compare \shortfigref{fig:sub_rel_mean_size_activity} to \shortfigref{fig:sub_mean_size_activity}).
For feedback,
as users receive better feedback over time,
users in \high activity \level
receive worse feedback in the beginning
and catch up later in their life (\shortfigref{fig:sub_rel_median_activity}).
These results are
natural
consequences of the trend
developing over time.
This suggests that the trends that we observe are robust over user life.

\begin{figure}[t]
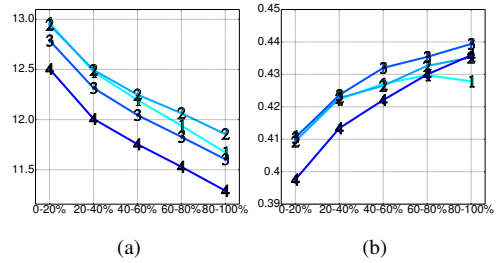

\centering
\subfigureplot{{plots/1107/sub_size/mean_log_size_relative_cmp_50_10_10_5}.pdf}{%
}{fig:sub_rel_mean_size_activity}{0.18}
\subfigureplot{{plots/1107/sub_median/median_relative_cmp_50_10_10_5}.pdf}{%
}{fig:sub_rel_median_activity}{0.18}
\caption{
Comparison of  different Reddit activity \levels from the \relative perspective.
(a): mean
$\log_2$(monthly number of posts).
(b):
fraction of posts that outperform the median value of feedback positivity
in the corresponding month and community.
\label{fig:sub_rel}}
\vspace{-0.15in}
\end{figure}

\para{\Absolute view for researchers in DBLP.}
In DBLP, authors span more conferences
per window over time (\shortfigref{fig:dblp_count_10}) in
an increasingly
scattered fashion (\shortfigref{fig:dblp_entropy_10}),
but in contrast to Reddit,
there is
saturation %
in the last two windows.
Perhaps this suggests that
as researchers
become very senior,
they publish more papers in
some favorite set of venues.
When a very small window size is considered ($w\mbox{=}5$),
the number of unique conferences and within-window entropy  first
increase and then decrease
(\shortfigref{fig:dblp_count_5} and \ref{fig:dblp_entropy_5}).
But, changing the window size
does \emph{not} affect our
central
observation in \shortfigref{fig:cumu_sub_count} that \major researchers are publishing in new conferences at a relatively consistent rate over the years.

\begin{figure}[t]
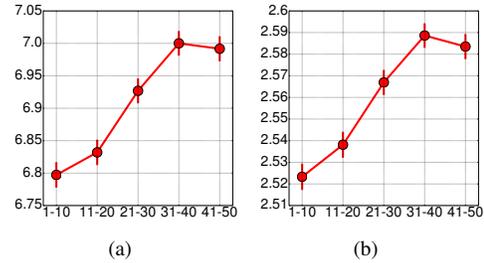

\centering
\subfigureplot{{plots/1012/dblp/count/absolute_all_50_10_10_5}.pdf}{%
}{fig:dblp_count_10}{0.18}
\subfigureplot{{plots/1012/dblp/entropy/absolute_all_50_10_10_5}.pdf}{%
}{fig:dblp_entropy_10}{0.18}

Window size of 10.

\vspace{-0.1in}
\manline

\subfigureplot{{plots/1012/dblp/count/absolute_all_50_5_5_10}.pdf}{%
}{fig:dblp_count_5}{0.18}
\subfigureplot{{plots/1012/dblp/entropy/absolute_all_50_5_5_10}.pdf}{%
}{fig:dblp_entropy_5}{0.18}

Window size of 5.
\caption{\Absolute view for researchers in DBLP.
(a,c): number of unique conferences per window. (b,d):
entropy of the conference publishing distribution per window.
\label{fig:dblp_window_10}}
\end{figure}

\newcommand{\finit}[2]{#1.}
\para{Acknowledgments.}
``Not all those who wander are lost'' (J. R. R. Tolkien).
We would have been lost without
\finit{C}{ristian} Danescu-Niculescu-Mizil,
\finit{J}{ack} Hessel,
\finit{A}{rzoo} Katiyar,
\finit{J}{on} Kleinberg,
\finit{B}{o} Pang,
\finit{F}{ilip} Radlinski,
\finit{A}{mit} Sharma,
\finit{K}{arthik} Sridharan,
\finit{A}{dith} Swaminathan,
\finit{Y}{isong} Yue,
the Cornell NLP seminar participants and the reviewers for their comments,
and Jason Baumgartner for redditanalytics.com.
This work was supported in part by NSF grant IIS-0910664 and a Google Research Grant.

\small
\bibliographystyle{abbrv}
\bibliography{ref}

\end{document}